\documentclass{emulateapj}

\slugcomment{Accepted by the Astrophysical Journal }

\shorttitle{PSR J2021+3651 with the {\em Fermi\/} LAT}
\shortauthors{Abdo et al.}

\begin{document}


\title{Pulsed Gamma-rays from PSR J2021+3651 with the {\em Fermi\/} Large Area 
Telescope}

\author{
A.~A.~Abdo\altaffilmark{1,2}, 
M.~Ackermann\altaffilmark{3}, 
M.~Ajello\altaffilmark{3}, 
W.~B.~Atwood\altaffilmark{4}, 
L.~Baldini\altaffilmark{5}, 
J.~Ballet\altaffilmark{6}, 
G.~Barbiellini\altaffilmark{7,8}, 
D.~Bastieri\altaffilmark{9,10}, 
M.~Battelino\altaffilmark{11}, 
B.~M.~Baughman\altaffilmark{12}
% ...
}
\altaffiltext{1}{National Research Council Research Associate}
\altaffiltext{2}{Space Science Division, Naval Research Laboratory, Washington, DC 20375}
\altaffiltext{3}{W. W. Hansen Experimental Physics Laboratory, Kavli Institute for Particle Astrophysics and Cosmology, Department of Physics and Stanford Linear Accelerator Center, Stanford University, Stanford, CA 94305}
\altaffiltext{4}{Santa Cruz Institute for Particle Physics, Department of Physics and Department of Astronomy and Astrophysics, University of California at Santa Cruz, Santa Cruz, CA 95064}
\altaffiltext{5}{Istituto Nazionale di Fisica Nucleare, Sezione di Pisa, I-56127 Pisa, Italy}
\altaffiltext{6}{Laboratoire AIM, CEA-IRFU/CNRS/Universit\'e Paris Diderot, Service d'Astrophysique, CEA Saclay, 91191 Gif sur Yvette, France}
\altaffiltext{7}{Istituto Nazionale di Fisica Nucleare, Sezione di Trieste, I-34127 Trieste, Italy}
\altaffiltext{8}{Dipartimento di Fisica, Universit\`a di Trieste, I-34127 Trieste, Italy}
\altaffiltext{9}{Istituto Nazionale di Fisica Nucleare, Sezione di Padova, I-35131 Padova, Italy}
\altaffiltext{10}{Dipartimento di Fisica ``G. Galilei", Universit\`a di Padova, I-35131 Padova, Italy}
\altaffiltext{11}{Department of Physics, Royal Institute of Technology (KTH), AlbaNova, SE-106 91 Stockholm, Sweden}
\altaffiltext{12}{Department of Physics, Center for Cosmology and Astro-Particle Physics, The Ohio State University, Columbus, OH 43210}
% ...

\begin{abstract}
We report the detection of pulsed gamma-rays from the young, spin-powered radio pulsar PSR J2021+3651 using data
acquired with the Large Area Telescope (LAT) on 
the {\em Fermi Gamma-ray Space Telescope\/} (formerly GLAST). 
The light curve consists of two narrow peaks of similar amplitude separated by $0.468 \pm 0.002$ in phase. The first
peak lags the maximum  of the 2 GHz radio pulse by $0.162 \pm 0.004 \pm 0.01$ in phase. 
The integral gamma-ray photon flux above 100 MeV is $(56\pm 3 \pm 11)\times 10^{-8}\,$cm$^{-2}$\,s$^{-1}$.
The photon spectrum is well-described by an exponentially cut-off power law of the form
${dF \over dE} = kE^{-\Gamma} e^{(-E/E_c)}$ where the energy $E$ is expressed in GeV.
The photon index is $\Gamma = 1.5 \pm 0.1 \pm 0.1$ and the
exponential cut-off is $E_c = 2.4 \pm 0.3 \pm 0.5$ GeV.
The first uncertainty is statistical and the second is systematic. 
The integral photon flux of the bridge is approximately 10\% of the pulsed emission, and
the upper limit on off-pulse gamma-ray emission from a putative pulsar wind nebula is $<10\%$ of the pulsed emission  
at the 95\% confidence level. 
Radio polarization measurements yield a rotation measure of $\mbox{RM} = 524\pm4$\,rad\,m$^{-2}$ but a 
poorly constrained magnetic geometry.
Re-analysis of {\em Chandra \/} data enhanced the significance of the weak X-ray pulsations, 
and the first peak is roughly phase-aligned with the first gamma-ray peak.
We discuss the emission region and beaming geometry based on the shape and spectrum of the 
gamma-ray light curve combined with
radio and X-ray measurements, and the implications for the pulsar distance. 
Gamma-ray emission from the polar cap region seems unlikely for this pulsar. 
\end{abstract}


\keywords{pulsars: general --- pulsars: individual (PSR J2021+3651) --- gamma-rays: observations }

\section{Introduction}
The Large Area Telescope (LAT) went into orbit on 2008 June 11
aboard the {\em Fermi Gamma-ray Space Telescope\/} \citep[formerly GLAST;][]{LATpaper}. 
Gamma-ray pulsations from the pulsar PSR J2021+3651 \citep{MR02} 
were detected during the first weeks of commissioning.
The LAT was built to address, along with other pressing questions 
in high energy astrophysics, the extent to which gamma-ray emission by  
pulsars is a rule, and thereby to better understand the mechanisms by which the kinetic energy of 
a rotating neutron star is transformed into intense beams of radiation.
The discovery of a pulsed gamma-ray signal from PSR J2021+3651 was reported using
{\em AGILE\/}, the ``Astro-rivelatore Gamma a Immagini LEggero'', by Halpern et al. (2008).

Following the detection of at least six high-energy gamma-ray pulsars by the Energetic Gamma 
Ray Experiment Telescope (EGRET) on the {\em Compton Gamma Ray Observatory\/} (see summary by 
Thompson 2004), observers sought positional coincidences of newly-discovered pulsars (e.g., Kramer et al. 
2003) with EGRET catalog sources (Hartman et al. 1999) or searched the EGRET error boxes for 
new radio pulsars (e.g., Crawford et al. 2006).

In one such search, Roberts et al. (2002) discovered PSR J2021+3651 with period $103.7$ ms within the GeV error box 
associated with 3EG J2021+3716 (Lamb \& Macomb 1997; Roberts et al. 2001). 
PSR J2021+3651 is young (characteristic timing age 17 kyr) and energetic 
($\dot E=3.38\times 10^{36}$ ergs s$^{-1}$). 
The NE2001 model (Cordes \& Lazio 2002) for the Galactic distribution of free electrons for 
this line of sight ($l = 75\fdg21,\, b = 0\fdg13$)
and the dispersion measure of $\mbox{DM} \approx 370$ pc cm$^{-3}$ suggest a distance of $D = 12$ kpc, with
a fractional uncertainty that can exceed 50\%. 
Subsequent investigations with the {\em Chandra X-ray Observatory\/}
revealed a pulsar wind nebula (PWN) and 
possible pulsations in X-rays (Hessels et al. 2004).
The torus of this ``Dragonfly'' PWN is clearly 
resolved, providing an estimate of the orientation of the pulsar's spin axis relative to the 
observer's line of sight of $85^\circ \pm 1^\circ$ (Van Etten et al. 2008).
All authors point out that such a large distance implies high
efficiency $\eta$ for the conversion of spin-down power into gamma-rays, becoming
unphysical ($\eta > 100\%$) for some beam scenarios, and explore different
ways to constrain the distance.  
The X-ray spectra are consistent with a distance of 2 to 4 kpc.
This paper adds new elements to the discussion: 
a more detailed gamma-ray light curve with phase-resolved spectroscopy,
a detailed comparison of the gamma-ray light curve with the predictions
of various models, and radio polarization measurements.

PSR J2021+3651 illustrates the importance of sustained monitoring of pulsar timing.  
It is one of the more than 200 pulsars with $\dot E > 10^{34}$ ergs s$^{-1}$
monitored by the program coordinated between 
the radio and X-ray timing community and the {\em Fermi\/} LAT team (Smith et al. 2008). 
PSR J2021+3651, like many young pulsars, exhibits timing noise. Not only does contemporaneous
timing allow accurate comparisons between radio and gamma-ray light curves,
but folding at a known period provides significantly greater sensitivity than searching for a periodicity.
McLaughlin \& Cordes (2004) detected significant periodicities for PSR
J2021+3651 in two of eight EGRET viewing periods (VPs) containing the
pulsar by extrapolating the timing solution and searching around a
range of period and period derivative. The light curve was consistent
between the two VPs and is similar to the LAT light curve.
However, they were unable to detect significant pulsations in the
other six VPs. PSR J2021+3651 was one of only two pulsars
detected by them in this manner due to the large number of trials
needed when extrapolating timing solutions back several years. The
ongoing pulsar monitoring for {\em Fermi\/} greatly reduces such problems.

\section{Observations}

The LAT is a pair-production telescope.  
It consists of tungsten foil and silicon microstrip 
converter/trackers (pair conversion and track measurement); hodoscopic cesium iodide 
calorimeters (energy measurement); plastic scintillator anticoincidence detectors 
(charged-particle rejection); and a programmable trigger and data acquisition system.  The LAT's 
excellent sensitivity stems from a large effective area ($\sim 8000$ cm$^2$) and superior angular resolution. 
The broad field of view ($2.4$ steradian) allows long exposures to the whole sky.
The LAT is sensitive to gamma-rays with energy $>20$ MeV. 
Verification of the on-orbit response is continuing but appears consistent with 
expectations\footnote{\url{http://www-glast.slac.stanford.edu/software/IS/glast\_lat\_performance.htm 
}.}.

Gamma-ray events recorded with the LAT have time\-stamps that derive from a GPS clock 
on the {\em Fermi\/} satellite. 
Ground tests using cosmic ray muons demonstrated that the LAT measures event times
with a precision significantly better than 1 $\mu$s. 
On orbit, satellite telemetry indicates comparable accuracy.
Transformation to the solar system barycenter and phase calculations were done 
with the {\em Fermi\/} LAT ``Science Tools'', 
shown to be accurate to better than a few $\mu$s for isolated pulsars (Smith et al. 2008). 
Degradation of the barycentered time resolution from uncertainty in {\em Fermi}'s position is
negligible. 
End-to-end performance of the timing systems was confirmed using the bright EGRET pulsars.
The LAT gamma-ray phases of 
these pulsars relative to the radio phases agree with previous measurements  
\citep[e.g.,][]{JMFBig3}. 

Data were acquired in two different {\em Fermi\/} observing modes. 
Between 2008 June 30 and August 3 the LAT was often pointed near the northern orbital
pole ($\alpha= $18h40m, $\delta = 60^\circ$,
or $l = 90^\circ,\, b = 25^\circ$, in mid-July).
Since then, {\em Fermi\/} has been scanning the entire 
sky every two orbits (approximately 3 hours). The data used here were acquired through 2008 November 15.
The ``diffuse'' event selection was used (``Pass 6'', version 1, see Atwood et al. 2009), leaving a background
due to charged cosmic rays comparable to or less than the extragalactic diffuse gamma-ray emission.
Gamma-rays with measured zenith angles greater than $105^\circ$ were excluded, due to the intense
gamma-ray emission from the Earth's limb caused by cosmic rays interacting in the atmosphere. 

PSR J2021+3651 is being observed by the timing consortium supporting {\em
Fermi\/} observations with the NRAO Green Bank Telescope (GBT), the NAIC
Arecibo telescope, and the Lovell telescope at Jodrell Bank.  The combined
usage of these observatories provides simultaneously good timing precision
and good sampling.  The  most precise timing measurements are acquired
at Arecibo and GBT, while the best sampling is obtained with GBT and
Jodrell Bank; the observational setup at the latter is described in
Hobbs et al. (2004). The WAPP spectrometer used at Arecibo is described in Dowd et al. (2000).
Here we use a rotational ephemeris based on GBT data.

The phase-connected ephemeris for PSR~J2021+3651 listed in Table 1,
contemporaneous with the {\em Fermi\/} observations, was derived
from 21 observations obtained between 2008 June 17 and November
15\footnote{This and other ephemerides used in {\em Fermi\/} results
will be available from the Fermi Science Support Center (FSSC) data
servers at http://fermi.gsfc.nasa.gov/ssc.}.  
The DE405 solar system ephemeris was used \cite{DE405}.
The pulsar was
observed at a center frequency of either 1950\,MHz or 1550\,MHz with
the GBT Pulsar Spigot (Kaplan et al.\ 2005), yielding total power
samples every $81.92\,\mu$s in each of 768 frequency channels over
a bandwidth of 600\,MHz.  Each observation lasted for 5 minutes,
from which we derived a time of arrival with typical uncertainty
of 0.2\,ms.  PSR~J2021+3651 exhibits rotational instability that
is significant over the span of 5 months.  We used the TEMPO timing
software\footnote{http://www.atnf.csiro.au/research/pulsar/tempo/}
and describe the pulsar rotation well by fitting for its frequency
and first two derivatives.  This timing solution had  small unmodeled
residual features ($\chi^2_\nu = 1.4$), with a post-fit rms of 0.2\,ms.
The DM is used to correct the time of arrival of the radio pulse to
infinite frequency for absolute phase comparison with the gamma-ray
profile.  Because the radio profiles of PSR~J2021+3651 are significantly
scattered at the frequencies that we use for timing, an ordinary TEMPO fit
biases the DM upwards from its true value. In order to correct for this,
we fit scattering models to the (assumed intrinsic) profile at 5\,GHz
until the resulting profiles matched the observed ones. 
The measured pulse broadening (scaled to 1 GHz) appears to be about a factor of two larger than
previously estimated by Hessels et al. (2004).
The corresponding
$\mbox{DM} = 367.5\pm1$\,pc\,cm$^{-3}$ is 0.5\% smaller than the value
without this correction, where the estimated uncertainty is dominated
by systematic effects. 
The resulting phase uncertainty after extrapolation from $1.95$\,GHz to infinite frequency is
$\pm 0.01$.

In addition to the Spigot observations, we observed the pulsar at 2000~MHz
using the new Green Bank Ultimate Pulsar Processing Instrument 
(GUPPI)\footnote{https://wikio.nrao.edu/bin/view/CICADA/GUPPiUsersGuide}
at the GBT for $1.3$ hours. 
GUPPI provides full-polarization
spectra in 2048 channels over 800~MHz of bandwidth every $40.96$ $\mu$s.  
The resulting data were polarization- and
flux-calibrated based on observations of a local pulsed noise signal and
the bright quasar J1445+0958. Analysis was performed using the
PSRCHIVE software package (Hotan et al. 2004).  

We re-analysed the $20.8$ ks {\em Chandra \/} ACIS-S continuous clocking data
from MJD 52682 described in Hessels et al. (2004), using the current version of {\em Chandra \/}
analysis software, and a timing solution built from near-contemporaneous radio
observations acquired with the Lovell telescope at Jodrell Bank.
The light curve is shown in Figure \ref{lightcurve}.
The deviation from a flat light curve is slightly more significant ($4.5\sigma$) than
that reported by Hessels et al. ($3.7\sigma$), has the same overall shape as their
Figure 3, but is shifted in phase. 
The first peak seems to be within $\approx 0.1$ in phase with the first gamma-ray peak
albeit with weak statistics.
Hessels et al. discuss the possibility of
a non-thermal component from the X-ray point source, and Van Etten et al.
measure this component's flux: the ratio of the X-ray non-thermal flux
to the gamma-ray flux is $10^4$, 
greater than that observed for Vela ($10^3$) or for Geminga ($10^2$, Bignami \& Caraveo 1996).  
The X-ray light curve also resembles those
of other established gamma-ray pulsars (Kaspi et al. 2006).

\section{Results}
The LAT orientation with respect to the celestial sphere has been
calibrated to a precision of $30^{''}$, using more than a dozen known
bright point sources. The {\em Chandra}-derived position of PSR J2021+3651 is $0\farcm62$ 
from the center of the gamma-ray 68\% containment contour, which has a $0\farcm75$ radius.
Figure \ref{counts} shows the distribution of counts for the region surrounding the pulsar.

\subsection{Light curve}
The top frame of Figure \ref{lightcurve} shows the phase histogram for the gamma-ray events with energies
$>100$ MeV, within an energy-dependent 68\% angular containment 
radius of the radio pulsar position.
There are two peaks: 
the first peak (P1) is the phase interval $0.13 < \phi < 0.20$, and P2 is $0.58 < \phi < 0.68$.
Both peaks are well-fit by a Lorentzian function, yielding
full widths at half-maximum of $0.021 \pm 0.002$ 
and $0.053 \pm 0.006$ rotations, respectively, separated by 
$\Delta\phi_{\gamma} = 0.468 \pm 0.002$. 
P1 lags the maximum of the 2 GHz radio peak shown in the bottom frame
by $\delta\phi_{\gamma-radio} = 0.162 \pm 0.004 \pm 0.01$ rotations of the neutron star.
The first uncertainty is statistical, the second is due to the uncertainty in the DM.
The intermediate frames show three energy bands: 100 MeV to 1 GeV; 1--3 GeV; and $>3$ GeV. 
P1 fades with increasing energy, whereas P2 persists.
The highest energy photon in this sample has 12 GeV, in P2.
The peak positions are stable with energy to within $0.01$ in phase.

\subsection{Spectrum}
The spectral shape, the cutoff energy, and the integral energy flux
are observables that can be compared with pulsar emission models.
PSR J2021+3651 is located in the Cygnus region where diffuse emission is
bright and there are poorly resolved neighboring sources (Fig. \ref{counts}). 
While most spectra in the 3rd EGRET catalog were well-modeled by a single
power-law spectrum, 3EG J2021+3716 was a clear exception.
A two power-law fit, with the break point fixed at 1 GeV,
gave a reduced $\chi^2=0.55$, compared to reduced $\chi^2 = 2.8$ for 
the simple power law (Bertsch et al. 2000; Reimer \& Bertsch 2001). 
The two spectral indices of $1.23 \pm 0.15$ 
and $3.39 \pm 0.36$ bracket the 3EG index of $1.86 \pm 0.10$.
The integral photon flux of $35\times 10^{-8}$\,cm$^{-2}$\,s$^{-1}$ thus obtained
is a bit more than half of the 3EG value, presumably because of
how neighboring sources were handled.
{\em AGILE\/} data confirm a spectral break near 2 GeV \citep{Agile}.

The different approaches used in analyses of LAT data to find the background contribution and to take into
account the direction- and energy-dependent instrument response are outlined in 
our recent analysis of LAT data for Vela (Abdo et al. 2009). One method estimates the background and
pulsar spectrum by maximizing the joint likelihood for the off-pulse data,
phases $0.73 < \phi <1.05$, and the pulsed data (`on-off').
Another models observed neighboring sources and the diffuse Galactic, extragalactic, and residual
charged backgrounds, again in a likelihood approach (`gtlike', provided with the {\em Fermi\/} science
tools).  
The `unfolding' method deconvolves the observed events from the instrument response. All three
methods give consistent results. 

Figure \ref{SpectrumAll} shows the result of the likelihood fit to the data assuming
a power-law spectrum with an exponential cutoff, 
$${dF \over dE} = kE^{-\Gamma} e^{(-E/E_c)^b}$$
where the energy $E$ is expressed in GeV and with $b=1$, using only the on-pulse data
($0.05 < \phi < 0.73$). The curves are the `gtlike' results, whereas `on-off' gives a
more faithful representation of the individual source data points, which are shown.
To reduce confusion from the diffuse background and neighboring sources, we fit only photons
above 200 MeV and extrapolate the result to 100 MeV.
The spectral parameters are listed in Table \ref{SpecTab}, including the integral
energy flux $h = (4.3 \pm 0.1) \times 10^{-10}\,{\rm ergs\,cm^{-2}\,s^{-1}}$ for $E>100$ MeV.
The Pearson $\chi^2$ value for the fit is $1.1$.
Applying the likelihood ratio test against a spectral model with $b$ left free,
there is no evidence to reject the simple exponential model, 
a super-exponential cutoff ($b=2$) is disfavored at the $3\sigma$ level, 
and a pure power-law ($b=0$) is strongly excluded at $13\sigma$.
%

\begin{figure}
\includegraphics[width=0.65\textwidth]{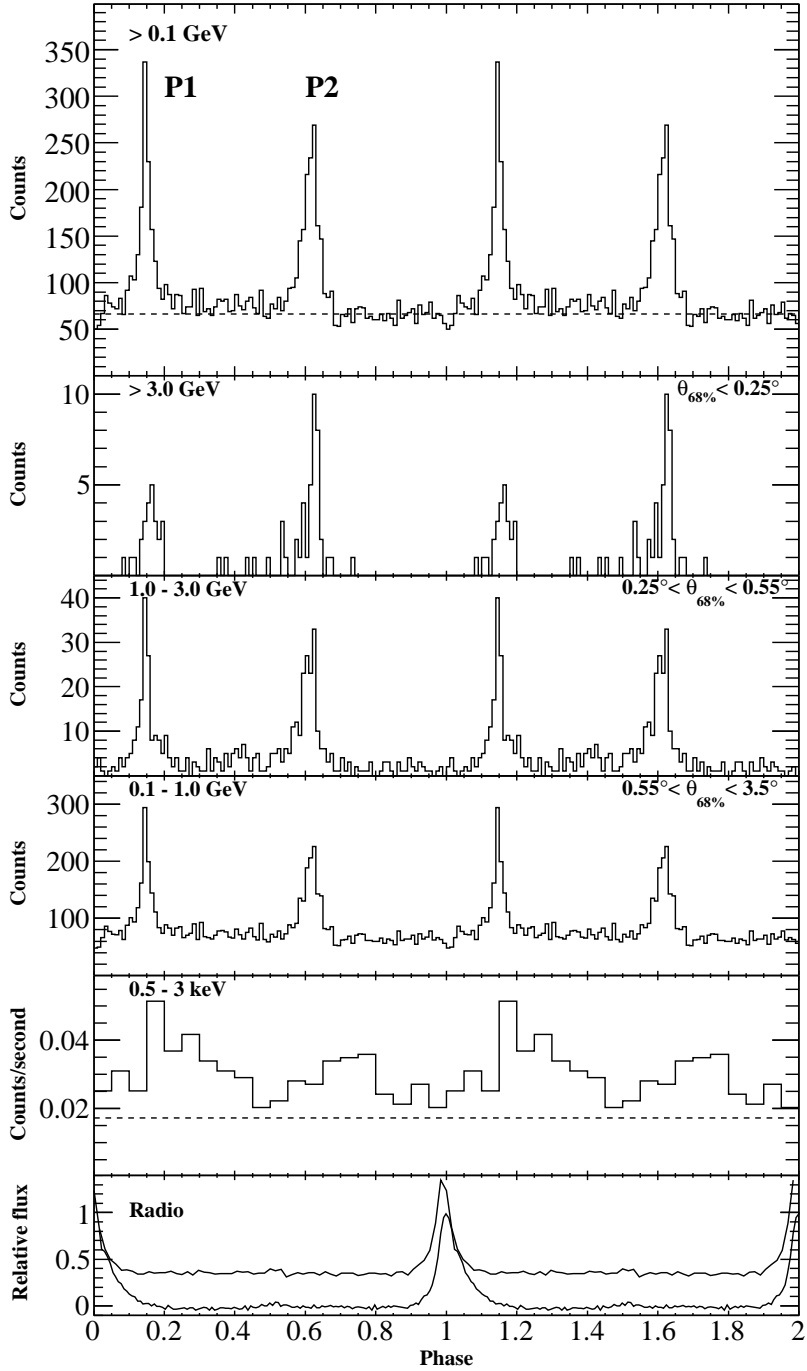}
\caption{ Top frame: Light curve for PSR J2021+3651, for gamma-rays with energy
$>100$ MeV within the energy-dependent 68\% containment radius $\theta_{68\%}$ 
of the pulsar position. 
Each bin is $0.01$ in phase, and 2 rotation cycles are shown. 
The horizontal dashed line shows the average number of counts
in the off-pulse phase interval.
Three following frames: Light curves in the three indicated energy ranges.
Second frame from bottom: Phase-aligned {\em Chandra \/} ACIS-S CC X-ray light curve, 
with the background rate shown by the horizontal dashed line.
Bottom frame:
The upper curve is the 1950 MHz radio profile obtained using the Pulsar Spigot at the Green Bank Telescope.
The lower curve is the
total intensity profile obtained at Arecibo using the WAPP spectrometers
with a 3.7\,hr integration time and 300\,MHz of bandwidth centered at 1500 MHz, vertically offset from
the GBT curve for clarity.  Both curves show evidence for an
interpulse at phase $0.5$ with amplitude $\sim 5\%$ of the main pulse.  
\label{lightcurve}}
\end{figure}

\begin{figure}
\includegraphics[width=0.6\textwidth]{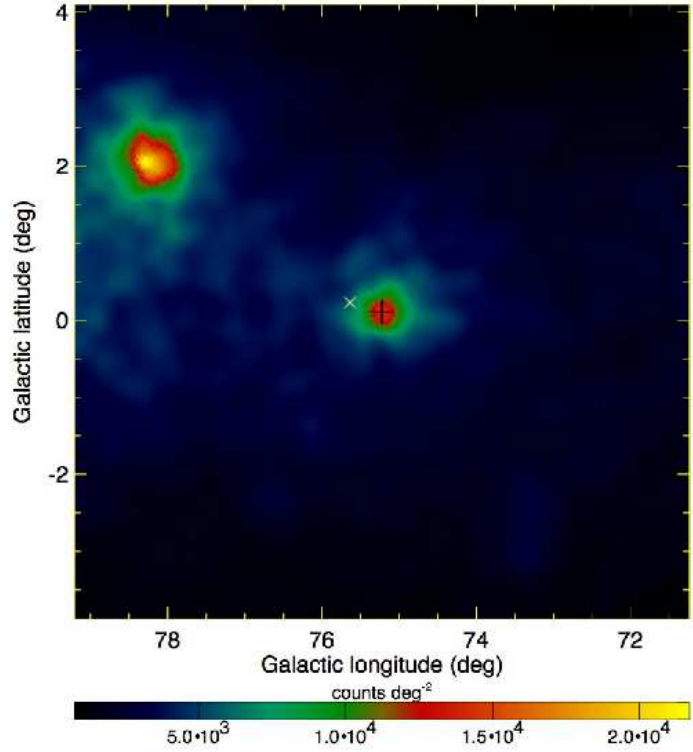}
\caption{Gamma-ray counts per square degree, with $E>100$ MeV,
centered at the {\em Chandra}-derived position of PSR J2021+3651.
Bin sizes vary such that statistical fluctuations are fixed to a signal-to-noise
ratio of 10.
The cross indicates the {\em Chandra}-derived position of PSR J2021+3651.
The ``X'' indicates the position of the open cluster Berkeley 87.
 The bright object at upper-left is 3EG J2020+4017. 
\label{counts}}
\end{figure}

\begin{figure}
\includegraphics[width=0.6\textwidth]{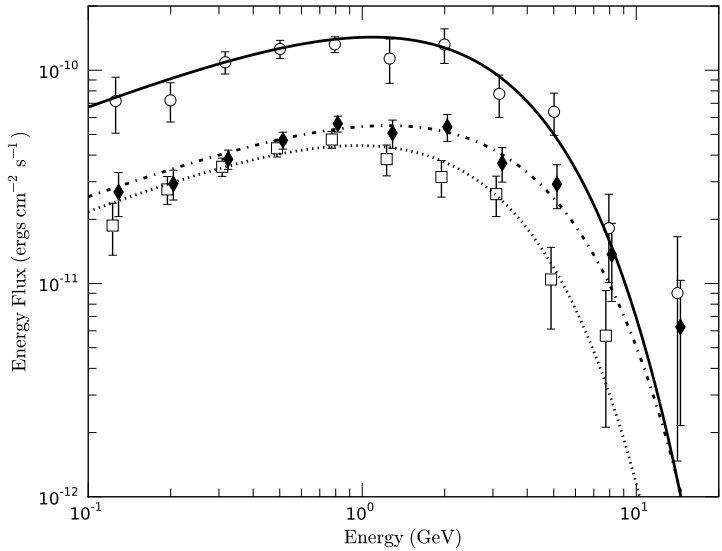}
\caption{Spectral energy distribution $E^2 dF/dE$ for PSR J2021+3651 as fit by ``gtlike'' assuming a power-law spectrum with
an exponential cutoff, for P1 (dotted), P2 (dot-dash), and Total pulse (solid).
The differential values as estimated by ``on$-$off'' are given for P1 (squares), P2 (diamonds),
and Total pulse (circles). The error bars are statistical only. 
For clarity, the points for P1 (P2) are plotted $2.5$\% lower (higher) in energy than
for Total pulse. 
\label{SpectrumAll}}
\end{figure}

We also measure the spectra of the two pulsar peaks individually.
Applying the same background model and again imposing $b=1$ yields the additional curves shown
in Figure \ref{SpectrumAll} and summarized in Table \ref{SpecTab}. P2 appears to
persist to higher energies in Figure \ref{lightcurve}, and the fit results formally
confirm this impression. However, in light of the systematic biases at play, the data
are also consistent with no significant spectral difference between the two pulses.
Bridge emission in the phase interval $0.26 < \phi < 0.54$ is apparent in Figure \ref{lightcurve}.
It exceeds the backgrounds from the diffuse Galactic emission and from
neighboring sources with $5 \sigma$ significance, with a photon flux that contributes roughly 10\%  
of the total on-pulse flux.
Finally, the off-pulse data show no excess above the background, allowing
us to place an upper limit on the flux of a putative gamma-ray PWN of
$ <10\%$ of the phase-averaged emission, at the 95\% confidence level.

The uncertainties in Table 2 and above are statistical. 
Two effects dominate the systematic biases:
modeling of the diffuse emission and neighboring sources
over the several degree radius dictated by the point spread function
at the lower energy bound ; 
and uncertainties in the energy-dependent effective area.
The latter is 
calculated using the test beam-verified Monte Carlo detector simulations (Baldini et al. 2007), 
and verified on-orbit using gamma-ray data from the Vela pulsar. 
For the LAT Vela spectral measurements (Abdo et al. 2009), the differences between
observed and expected on-orbit gamma-ray efficiencies led to a
uncertainty on the integral energy flux of $\pm \delta h/h = 20$\%.
Since then, the differences have been found to arise from 
charge in the silicon tracker deposited by cosmic rays in a time window 
around the gamma-ray event.
Gamma-ray event reconstruction and selection efficiencies below several hundred MeV are {\em smaller}
than predicted, and the fluxes reported here will increase for future 
analyses taking the effect into account. 
The potential bias in the cutoff energy
is of order $\pm 0.5$ GeV and that of the spectral index is $\delta\Gamma \approx 0.1$.
The conclusion that the spectral shape for PSR J2021+3651
is most consistent with $b=1$ is unaffected by these issues.

\section{Discussion}

PSR J2021+3651 is among the first gamma-ray pulsars to be studied using {\em Fermi}.
The gamma-ray observations, combined with X-ray images and spectra as well as with
radio information such as polarization, allow stricter comparison with
models than was previously possible. Improved knowledge of beam geometries will
in turn aid interpretation of the number counts of radio-loud and radio-quiet gamma-ray
pulsars that {\em Fermi\/} sees. 

Emission models fall into two classes: Polar cap (PC) scenarios (Daugherty \& Harding 1996),
which place the emission very close to the star surface, and outer magnetosphere models.
Of the latter,
the outer gap (OG) picture (Romani \& Yadigaroglu 1995;  
Cheng et al. 2000; Hirotani 2005) ascribes
the double pulse to emission at the boundaries of a single magnetic
pole, while the two pole caustic model (Dyks \& Rudak 2003), a physical realization of
which may be the modern
version of the slot-gap (SG) picture (Muslimov \& Harding 2004; Harding et al. 2008), 
assigns the two pulses to the trailing boundaries of separate magnetic poles.

Gamma-rays created at the polar caps interact with  
the intense magnetic fields near the neutron star surface, 
resulting in superexponential spectral cut-offs below a few GeV,
while OG models predict simple exponential cutoffs. 
LAT confirms and refines the spectral break measured with EGRET and also
seen with {\em AGILE\/} by Halpern et al. (2008). The observed emission beyond 10 GeV and absence of a sharp cut-off indicate
outer magnetosphere processes.

From the phase separation of the peaks in the gamma-ray light curve $\Delta\phi_\gamma=0.468$,
lag from radio $\Delta\phi_{\gamma-radio}=0.16$, and X-ray torus-derived
viewing angle $\zeta = 85\pm 1^\circ$, we have bounds on the
pulsar geometry that constrain the possible emission models.  
For the PC model $\Delta\phi_\gamma=0.5$ is natural for large $\zeta$ and 
magnetic inclination $\alpha$; one would expect to observe both
radio pulses as well, since the observer must view the emission at a
small angle $\beta = \zeta - \alpha$ to the magnetic pole to intersect the small PC
beam. Indeed, Figure \ref{lightcurve} shows a faint radio interpulse. On the other hand both outer
magnetosphere models can produce two narrow gamma-ray pulses at the
observed separation for $\zeta=85^\circ$. 
Narrow gamma-ray pulses arise naturally in outer
magnetosphere models, where the peaks are formed by caustics. 
For OG and SG, a magnetic inclination of $\alpha \approx 70^\circ$ is inferred. For this
$\alpha$, the second radio pole may be
faint (``grazing'') or absent.

\begin{figure}
\includegraphics[width=0.4\textwidth,angle=-90]{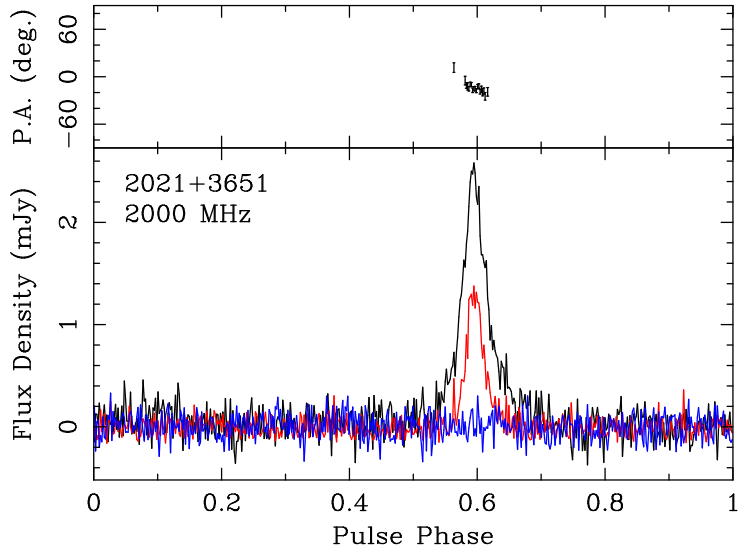}
\caption{ Bottom:  Polarization- and flux-calibrated profile
for PSR~J2021+3651 obtained with the GUPPI spectrometer at the GBT. 
The black trace corresponds to total intensity, the red to linear polarization,
and the blue to circular polarization (greatest to least intensity at
$\phi \approx 0.6$, respectively).  
The phase reference is different from Fig. \ref{lightcurve}. 
The faint interpulse is visible near phase $\approx 0.1$.
Top:
Position angle of linear polarization, corrected to the
pulsar frame, accounting for the Faraday rotation implied by the large
rotation measure of $\mbox{RM} = 524\pm4$\,rad\,m$^{-2}$.}
\label{polarz}
\end{figure}

An additional radio observation supports the view that PSR J2021+3651
is a nearly orthogonal rotator, that is, has a large magnetic inclination. 
Figure \ref{polarz} shows polarization data 
from which we determined the rotation measure (RM) of J2021+3651 to be
524$\pm$4~rad~m$^{-2}$, and measured the polarized pulse profile.
The data are insufficient to constrain the parameters of 
the Rotating Vector Model \cite[RVM,][]{RVM}:
they are compatible with $\alpha \approx 70^\circ$ found above.

The large lag from the radio pulse is a challenge for
simple versions of all models using a vacuum dipole magnetosphere.
For the PC model we expect $\Delta\phi_{\gamma-radio}=0$, in strong
disagreement. Both the OG and SG models predict
$\Delta\phi_{\gamma-radio}= 0.05-0.1$, for radio emission at the star surface.
However, recent work (Johnston \& Weisberg 2006; Karastergiou \& Johnston 2007) suggests
that radio emission from young, Vela-type pulsars occurs over a narrow range
of altitudes below roughly 100 times the radius of the neutron star, whereas
for their older brethren, it extends to lower altitudes. At high altitudes aberration
shifts the radio pulse forward and widens the radio cone. If only the leading
edge of the cone contributes to the narrow radio pulse, then the observed lag is
easily achieved.
%

The emission scenarios predict different relations between the
gamma-ray luminosity $L_\gamma$ and the spin-down power $\dot E$,
making the efficiency $\eta = L_\gamma/\dot E$ an additional discriminating
observable, especially if applied to a large sample of gamma-ray pulsars.
To obtain $L_\gamma$, the pulsar's distance $D$ must be known, and the observed integral
gamma-ray energy flux $h$ must be extrapolated to the full sky, that is, some model
of the beam shapes must be applied. In the past, the lack of geometric constraints
for the small number of known gamma-ray pulsars led to the convention of simply
assuming the gamma-ray beam swept out a 1 sr solid angle, from which
$L_\gamma = hD^2$; such a narrow beam is appropriate to near-surface polar cap emission.
To better exploit the available data
we define, following Watters et al. (2008),  a correction factor $f_\Omega$ along the Earth line-of-sight $\zeta_E$ as
\begin{equation}
 f_\Omega(\alpha,\zeta_E) = 
{\int F_\gamma(\alpha;\zeta,\phi)\sin(\zeta){\rm d}\zeta {\rm d}\phi \over 2\int F_\gamma(\alpha;\zeta_E,\phi){\rm d}\phi}
\end{equation}
such that
\begin{equation}
 L_\gamma = 4\pi f_\Omega h D^2.
\end{equation}
$F_\gamma(\alpha;\zeta,\phi)$ is the gamma-ray energy flux
as a function of $\zeta$ and the pulsar rotation phase $\phi$. 
In the ratio $f_\Omega$, the numerator
is the total emission over the full sky, and the denominator is the phase-averaged
flux for the light curve seen from Earth. 
A 1 sr sky coverage corresponds to
$f_\Omega = {1 \over 4\pi} = 0.08$ and isotropic emission gives $f_\Omega = 1$. 
Note that
$f_\Omega>1$ is possible for beams that are narrow in $\phi$, extended in $\zeta$,
and/or have average intensity exceeding the value sampled at $\zeta_E$. 

Polar cap models tend to have $f_\Omega \le 0.1$, while for the fan beams of
outer magnetosphere models $f_\Omega$ is much larger. For the OG model we estimate 
$f_\Omega\approx 1.05$ while the SG model has
$f_\Omega\approx 1.1$ for the observed $\zeta$ (Watters et al. 2008).

The distance to PSR J2021+3651 is intriguing. 
The NE2001 electron density model assigns a distance of $12^{+\infty}_{-2.7}$\,kpc
to the large DM of the pulsar, but also greatly underestimates
the measured scattering timescale, casting doubt on this inferred distance.
Van Etten et al. (2008) place the pulsar at $D=$ 2 to 4 kpc,
based on the PWN properties and the neutron star thermal emission.
The large positive RM presented here is consistent with a distance at or beyond 4\,kpc:
of the 9 pulsars within $\sim 10^\circ$ of this line-of-sight with 
measured RMs in the ATNF database\footnote{http://www.atnf.csiro.au/research/pulsar/psrcat/}
\citep{Manch}, those with $D < 5$ kpc have negative RMs, while
four others with $D \approx 6$ to 8 kpc have positive RM $< 150$ rad m$^{-2}$ and 
DM from 130 to 240 pc cm$^{-3}$. 

For a neutron star moment of inertia of $10^{45}$ g cm$^2$ we obtain
\begin{equation}
\eta = L_\gamma/ \dot E = 0.25 f_\Omega (D/4\,{\rm kpc})^2.
\end{equation}
High-altitude models are preferred by the pulse and spectral shapes.
A 25\% efficiency is amongst the largest for all known gamma-ray pulsars,
taking the $f_\Omega$ values as in Table 1 of Watters et al. (2008).
The nominal DM distance imposes small $f_\Omega$, i.e., polar cap
emission, whereas the smaller distances imposed by the high-altitude models
leave unexplained the large observed electron column density along the line of sight
to the Dragonfly nebula.
 
The open cluster Berkeley 87 is $0.5^\circ$ from PSR J2021+3651.
Prior to the pulsed gamma-ray detections, it was suspected to be a proton 
accelerator that could explain 3EG J2021+3716 and/or 3EG J2016+3657, and
was searched for TeV emission \citep[see e.g.][]{Hegra}.
The LAT localization clearly refutes Berkeley 87 as the dominant gamma-ray emitter
in this direction, even if the off-pulse emission upper limit is near the 
intensity predicted by Bednarek (2007), leaving the door open for future
explorations. The detailed maps in Schneider et al. (2007) include the position of 
PSR J2021+3651 and their re-examination may allow part of the large free electron 
column density to be accounted for. 


\section{Conclusions}
PSR J2021+3651 was detected in gamma-rays in LAT data taken during {\em Fermi\/} instrument commissioning,
with more data
 accumulated during the first few months of the all-sky survey.
The extensive radio pulsar timing being performed for the {\em Fermi\/} mission
facilitated the detection and enhanced the quality of the resulting light curves as
well as their interpretation.
Along with the discovery of a pulsed signal in the gamma-ray source CTA 1 
identifying a previously mysterious source (Abdo et al. 2008), 
this is a good example of the remarkable capability of the LAT to identify Galactic sources.
Of the 171 unidentified sources in the 3rd EGRET catalog, some 30
are Galactic ($|b|<3^\circ$) and have steady fluxes as measured by EGRET.
The identification of 3EG J2021+3716 suggests that many of these other sources could also
be pulsars. 

The high-resolution gamma-ray light curve, the faint radio interpulse, and the polarization data, 
together with earlier X-ray images of the torus and jet of the surrounding pulsar wind nebula, 
allow comparisons with different pulsar models. 
The rotation measure adds an argument in favor of an intermediate pulsar distance.
Re-analysis of {\em Chandra} X-ray data yields an improved light curve.
Phase-resolved spectral measurements show that both peaks cut
off exponentially near 2 GeV. Gamma-ray emission from the polar cap is the least 
plausible explanation at present, 
even if the outer magnetosphere models imply large gamma-ray
efficiencies if PSR J2021+3651 is indeed more distant than a few kpc.

\acknowledgments
The $Fermi$ LAT Collaboration acknowledges generous ongoing support from a number of agencies and institutes
that have supported both the development and the operation of the LAT as well as scientific data analysis. 
These include the National Aeronautics and Space Administration and the Department of Energy in the United
States, the Commissariat \`a l'Energie Atomique and the Centre National de la Recherche Scientifique / Institut
National de Physique Nucl\'eaire et de Physique des Particules in France, the Agenzia Spaziale Italiana and the
Istituto Nazionale di Fisica Nucleare in Italy, the Ministry of Education, Culture, Sports, Science and
Technology (MEXT), High Energy Accelerator Research Organization (KEK) and Japan Aerospace Exploration Agency
(JAXA) in Japan, and the K.~A. Wallenberg Foundation, the Swedish Research Council and the Swedish National
Space Board in Sweden.

Additional support for science analysis during the operations phase from the following agencies is also
gratefully acknowledged: the Istituto Nazionale di Astrofisica in Italy and the K.~A. Wallenberg Foundation in
Sweden for providing a grant in support of a Royal Swedish Academy of Sciences Research fellowship for JC.

The Green Bank Telescope is operated by the National Radio Astronomy Observatory, a
facility of the National Science Foundation operated under cooperative
agreement by Associated Universities, Inc. The Arecibo Observatory is part of the National Astronomy and Ionosphere 
Center (NAIC), a national research center operated by Cornell University 
under a cooperative agreement with the National Science Foundation.




\begin{deluxetable}{ll}
\tablewidth{0.65\linewidth}
\tablecaption{Radio ephemeris of PSR J2021+3651}
\startdata
\hline
Pulsar name \dotfill & J2021+3651 \\
Right ascension (J2000)\tablenotemark{a} \dotfill & 20:21:05.46 \\ 
Declination (J2000)\tablenotemark{a} \dotfill & 36:51:04.8 \\ 
Pulse frequency (s$^{-1}$)\dotfill & 9.6393948581(3) \\ 
Frequency derivative (s$^{-2}$)\dotfill & $-$8.89419(6)$\times 10^{-12}$ \\ 
Frequency 2nd derivative (s$^{-3}$)\dotfill & 1.09(5)$\times 10^{-21}$ \\
Epoch of fit (MJD [TDB])\dotfill & 54710.0 \\
Epoch of zero phase reference (MJD [TDB])\tablenotemark{b} \dotfill & 54715.22791423678 \\ 
Range of fit (MJD) \dotfill & 54634.18--54785.9 \\
Dispersion measure (pc cm$^{-3}$)\tablenotemark{c}\dotfill & $367.5 \pm 1$  
\enddata
\tablecomments{The digits in parentheses are the nominal $1 \sigma$
TEMPO uncertainties. }
\tablenotetext{a}{Celestial coordinates are from {\em Chandra\/}
observations (Hessels et al. 2004). }
\tablenotetext{b}{This is TEMPO's ``TZRMJD'' extrapolated
to the solar system barycenter at infinite frequency. }
\tablenotetext{c}{See \S~2 for details of this measurement. }
\end{deluxetable}

\begin{deluxetable}{lccccr}
\tablecaption{\label{SpecTab} Spectral results for PSR J2021+3651}
\startdata
\hline
Phase interval & $\Delta \phi$    &    $F$\tablenotemark{a}       &     $h$\tablenotemark{b}      &    $\Gamma$   &     $E_c$\tablenotemark{c}   \\ 
               &                  & ($10^{-8}\rm{cm}^{-2}{\rm s}^{-1}$) 
	       & ($10^{-10}\rm{ergs}\,\rm{cm}^{-2}{\rm s}^{-1}$) &               &    (GeV)    \\ \hline
P1             & $0.13$ to $0.20$ & $18 \pm 1$ & $1.3 \pm 0.1$ & $1.5 \pm 0.1$ &$1.9 \pm 0.3$ \\
P2             & $0.58$ to $0.68$ & $21 \pm 1$ & $1.7 \pm 0.1$ & $1.5 \pm 0.1$ &$2.8 \pm 0.5$ \\ 
Bridge         & $0.26$ to $0.54$ & $\approx 10$\% of Total pulse&              &               &              \\ 
Total pulse    & $0.05$ to $0.73$ & $56 \pm 3$ & $4.3 \pm 0.1$ & $1.5 \pm 0.1$ &$2.4 \pm 0.3$ \\ 
Off pulse      & $0.73$ to $1.05$ & $<10$\% of Total pulse &          &               &              
\enddata
\tablenotetext{a}{Integral photon flux ($> 100$ MeV). }
\tablenotetext{b}{Integral energy flux ($> 100$ MeV). Flux systematic biases could lead to 20\% increases, see text.}
\tablenotetext{c}{Energy of an exponential cut-off to a power-law spectrum
with index $\Gamma$. }
\end{deluxetable}

\end{document}